\documentclass[11pt]{article}

\catcode`\@=11

\global\arraycolsep=2pt
\usepackage{amsbsy,amssymb,latexsym,amsfonts,amsmath}
\usepackage{graphicx,color}

\begin{document}
\begin{flushright}
\parbox{4.2cm}
{CALT 68-2890}
\end{flushright}

\vspace*{0.7cm}

\begin{center}
{ \Large Is boundary conformal in CFT?}
\vspace*{1.5cm}\\
{Yu Nakayama}
\end{center}
\vspace*{1.0cm}
\begin{center}
{\it California Institute of Technology,  \\ 
452-48, Pasadena, California 91125, USA}
\vspace{3.8cm}
\end{center}

\begin{abstract}
We discuss boundary conditions for conformal field theories that preserve the boundary Poincar\'e invariance. As in the bulk field theories, a question arises whether boundary scale invariance leads to boundary conformal invariance. With unitarity, Cardy's condition of vanishing momentum flow is necessary for the boundary conformal invariance, but it is not sufficient in general. We show both a proof and a counterexample of the enhancement of boundary conformal invariance in $(1+1)$ dimension, which depends on the extra assumption we make. In $(1+2)$ dimension, Cardy's condition is shown to be sufficient. In higher dimensions, we give a perturbative argument in favor of the enhancement based on the boundary $g$-theorem. With the help of the holographic dual recently proposed, we show a holographic proof of the boundary conformal invariance under the assumption of the boundary strict null energy condition, which also gives a sufficient condition for the strong boundary $g$-theorem.
\end{abstract}

\thispagestyle{empty} 

\setcounter{page}{0}

\newpage

\section{Introduction}
Various defects play important roles in physics. Wilson loops and 't Hooft loops are fundamental gauge invariant observables in gauge theories, and D-branes play significant roles in string theory and beyond. We may observe cosmic strings in the sky while vortices and domain walls are ubiquitous in condensed matter physics.

In this paper, we would like to study the renormalization group properties of the codimention one boundaries in conformal field theories. Understanding of the renormalization group properties of the boundary conditions and the requirement of the boundary conformal invariance in $(1+1)$ dimension is equivalent to understanding of the dynamics of the D-branes in string theory. It is curious to note that although the importance of bulk conformal invariance had been long appreciated in condensed matter physics, the importance of the boundary critical phenomena was emphasized much later as was the realization of the importance of D-branes.

Cardy proposed a boundary condition \cite{Cardy:1984bb} which is compatible with the conformal invariance of the boundary. Cardy's condition demands that the momentum flow vanishes at the boundary: $T_{ri}=0 $, where $r$ is the normal direction of the boundary. In $(1+1)$ dimensional conformal field theories with the Virasoro symmetry, the condition leads to the so-called Ishibashi-states \cite{Ishibashi:1988kg} that are building block of the perturbative construction of D-branes in worldsheet string theory.\footnote{Often, the integrality of the open string spectrum constructed out of the D-brane boundary state is also called Cardy's condition in $(1+1)$ dimension. We have little to say about it in this paper.}

We would like to argue that although Cardy's condition in higher dimensions is necessary for the boundary conformal invariance with unitarity, generally it is not sufficient. Even the necessity can be lost when we abandon the unitarity of the boundary conformal field theories. For better understanding of the general features and possible swampland in quantum field theories, we would like to understand the condition as precisely as possible. This is one aim of this paper.

At the same time, we would like to propose a bold conjecture that the scale invariant boundary condition in conformal field theories are all conformal invariant. This cannot be true with no assumptions, but as in the bulk case, the additional assumption of Poincar\'e invariance (in particular causality), unitarity, existence of the energy-momentum tensor, and so on may lead to the enhancement of the symmetry. At least, we do not know an immediate counterexample for the conjecture.

In the bulk quantum field theories, the discussions over the scale invariance and conformal invariance have a relatively long history. In late 80's, it was established that the scale invariance implies conformal invariance in $(1+1)$ dimension \cite{Zamolodchikov:1986gt}\cite{Polchinski:1987dy}\cite{Mack1}. In $(1+3)$ dimension, it was realized that the Wess-Zumino consistency condition of the local renormalization group gives a perturbative proof of the equivalence \cite{Jack:1990eb}\cite{Osborn:1991gm}.\footnote{When we mean conformal invariance or scale invariance, we always assume that they are not spontaneously broken in the paper.} Recently, it was freshly revisited by using the dilaton effective action approach with the same conclusions \cite{Luty:2012ww}. After some debates, the counterexample proposed in \cite{Fortin:2012cq}\cite{Fortin:2012hn} actually turns out to be conformal invariant, and the results are perfectly consistent with the original argument based on the Wess-Zumino consistency condition (for clarification see Appendix of \cite{Nakayama:2012nd}). During our discussions for the enhancement of the boundary conformal invariance, we would like to comment on the similar subtleties.

Although we cannot make a definite statement in higher dimensions $(d>3)$, we will present an argument in favor of the conjecture from holography. We will generalize the holographic setup for the boundary conformal field theories proposed in \cite{Takayanagi:2011zk}\cite{Fujita:2011fp}, and discuss whether the scale invariant but non-conformal boundary conditions are possible for the bulk conformal field theories realized by $AdS_{d+1}$ bulk geometry. Under a certain energy condition, which is supposedly a sufficient condition for the unitarity in the bulk, we can show that the scale invariant boundary condition leads to the conformal invariant boundary condition.

The organization of the paper is as follows. In section 2, we review the boundary conditions for the conformal field theories that preserve the Poincar\'e invariance in tangential directions. We clarify the distinction between boundary scale invariance and boundary conformal invariance. In section 3, we study under which condition, the boundary scale invariance implies boundary conformal invariance. We pay special attention to the generalization of Cardy's condition in higher dimensions. We show both a proof and a counterexample in $(1+1)$ dimension, which depends on the extra assumption we make. In $(1+2)$ dimension, Cardy's condition is shown to be sufficient. In higher dimensions, we give a perturbative argument in favor of the enhancement based on the boundary $g$-theorem. In section 4, we try to understand the mechanism of the enhancement of the boundary conformal invariance based on the holographic approach. We show a holographic proof of the enhancement of the boundary conformal invariance under the assumption of the boundary strict null energy condition, which also gives a sufficient condition for the strong boundary $g$-theorem. In section 5, we conclude the paper with future directions.

\section{Boundary conditions for CFT}
Let us consider the boundary conditions for $d$-dimensional conformal field theories in Minkowski space-time that preserve the Poincar\'e invariance in tangential directions \cite{McAvity:1993ue}\cite{McAvity:1995zd}. We put the boundary at $r=0$, where $r$ is one of the spatial Cartesian coordinate of $\mathbf{R}^{1,d-1}$, and impose the Poincar\'e invariance on the $(x_0,x_1, \cdots, x_{d-2})$ plane.\footnote{From here on, we use $\mu = 0,1,2\cdots, {d-2},r$ and $i=0,1,2,\cdots, d-2$ when we study the field theories in $d$ dimension. }  We will exclusively study the case when the bulk theory is conformal invariant.\footnote{This is motivated by the argument for the enhancement of conformal invariance from scale invariance in bulk quantum field theories, which is non-perturbatively proved in $(1+1)$ dimension non-perturbatively, and perturbatively proved in $(1+3)$ dimension under certain reasonable assumptions such as unitarity.} Since we will discuss the enhancement of the boundary conformal invariance from the boundary scale invariance, it is less sensible to ask the question when the bulk does not possess the conformal invariance.

Under the Noether assumption, which dictates that the symmetries of the theory must be realized by local currents, the bulk conformal invariance implies that the theory must possess the bulk traceless-symmetric-conserved energy-momentum tensor $T_{\mu\nu}$ such that
\begin{align}
\partial^\mu T_{\mu\nu} = T^{\mu}_{\ \mu} =T_{\mu\nu} - T_{\nu\mu} = 0 \ .
\end{align}
Although in principle we can realize symmetries without Noether currents because only charges are necessary,  we always assume it is the case in the following discussions. 

We have to provide a boundary condition for the bulk energy-momentum tensor at $r=0$ that is compatible with the symmetry we would like to preserve. If we would like to preserve the Poincar\'e invariance on the boundary plane, the Noether theorem demands
\begin{align}
T_{r i} (r=0) = \partial^j \tau_{ji} \ , \label{btc}
\end{align}
where $\tau_{ij}$ is a symmetric tensor localized at the boundary $r=0$. We call $\tau_{ij}$ ``boundary energy-momentum tensor". There are no other boundary conditions on $T_{\mu\nu}$ that are required by the symmetries (even if we include the boundary conformal invariance). We will discuss the geometrical origin of the boundary energy-momentum tensor in \eqref{varim}.

Generically speaking, energy-momentum tensor is not unique. In the bulk, we have partially fixed the non-uniqueness by demanding the traceless condition, which makes the bulk conformal invariance manifest. The remaining ambiguity corresponds to the ambiguity hidden in the flat space-time action for different realization of the conformal symmetries (e.g. the background charge of the linear dilaton theory in $(1+1)$ dimension). 
Even with the fixed bulk energy-momentum tensor, the boundary energy-momentum tensor that satisfies \eqref{btc} is not unique. It is subject to the improvement 
\begin{align}
\tau_{ij} \to \tau_{ij} + \partial^l \partial^m r_{imjl} \ , \label{ambig}
\end{align}
where $r_{ijml}$ has the same symmetry of the Riemann tensor (in particular $r_{i[jkl]} = 0$)
because it does not affect the momentum flow condition \eqref{btc}.

We can introduce further requirements for the boundary condition. At the fixed point (trajectory) of the renormalization group flow, it is sensible to require the boundary scale invariance. It leads to the condition that the trace of the boundary energy-momentum tensor above introduced is given by the divergence of a certain current 
\begin{align}
\tau^{i}_{ \ i} = \partial^i j_i \ . \label{momentumflow}
\end{align}
With the bulk analogy in mind \cite{Coleman:1970je}, we would like to call $j_i$ the boundary virial current. We emphasize that the boundary energy-momentum tensor $\tau_{ij}$ and the boundary virial current $j_i$ are not the restriction of the bulk energy-momentum tensor and the bulk virial current on the boundary plane. 

In contrast, the kinematical condition from the Noether assumption for the boundary conformal invariance is that the boundary energy-momentum tensor can be improved to be traceless. 
The necessary and sufficient condition is that the boundary virial current is expressible as
\begin{align}
j_i &= \partial^j l_{ji}  \ \ ( d>3) \cr
    &=  \partial_i l     \ \ (d = 2,3) \ ,
\end{align}
where $l_{ij}$ and $l$ are local operators on the boundary.
Then we can improve the boundary energy-momentum tensor $\tilde{\tau}_{ij}$ by using the ambiguity \eqref{ambig} so that it is traceless.\footnote{When $d=2$, the situation is somewhat different. Instead of improving the energy-momentum tensor $\tau_{tt}$, which we cannot do, we simply add $-2tj_t + 2l$ to the special conformal generator $ t^2\tau_{tt} + \int dr K_t$.}
\begin{align}
\tilde{\tau}^{i}_{\ i} = \tau^{i}_{\ i} - \partial^i\partial^j l_{ij} = 0 \ . \label{traceless}
\end{align}
The boundary energy-momentum tensor might be still ambiguous after imposing the manifest boundary conformal invariance because we could improve it with $\partial^l \partial^m w_{imjl}$ with the tensor $w_{imjl}$ whose tensor structure is the same as the Weyl tensor. 

The discussions in this section have been purely kinematical. It must be true in any bulk conformal field theories with no constraint from unitarity or reflection positivity in Euclidean signature as long as the Noether assumption is satisfied. Once the boundary conformal invariance is achieved, one can conformally map our plane boundary $r=0$ to any conformally equivalent space-time (e.g. $d$-dimensional ball with $S_{d-1}$ boundary, or $AdS_d$ with $AdS_{d-1}$ boundary etc). 

Finally, we would like to briefly mention the general curved background formulation of the boundary condition for bulk conformal field theories \cite{McAvity:1993ue}. 
We start from the effective action 
\begin{align}
e^{-S_{eff}} = \int [D\phi] e^{-S_0[\phi]}
\end{align}
for fields $\phi$ on generic curved background $M$ with suitable boundary conditions on $ Q = \partial M$. The variation of the metric gives 
\begin{align}
\delta S &= \int_{\mathcal{M}} d^dx \sqrt{g} \frac{1}{2}\delta g^{\mu\nu} \langle T_{\mu\nu} \rangle \cr
&- \int_{Q} d^{d-1}x\sqrt{\gamma} \delta \gamma^{ij} \langle \tau_{ij} \rangle + \delta n^\mu \langle \lambda_{\mu} \rangle + \cdots  \ . \label{varim}
\end{align}
where $\delta n^\mu$ is the induced variation of the normal vector of the boundary, and we have neglected the extrinsic curvature term and further higher derivative terms at boundary. We can see that $\tau_{ij}$ is a response to the variation of the boundary metric $\gamma_{ij}$.

The invariance under the diffeomorphism $\delta g_{\mu\nu} = D_\mu v_\nu + D_\nu v_\mu$ gives the conservation equation
\begin{align}
D^\mu T_{\mu\nu}  = \lambda^\mu = 0 \cr
\end{align}
and the momentum flow condition
\begin{align}
T_{ri} = \hat{D}^j \tau_{ji} \label{conserve}
\end{align}
up to extrinsic curvature corrections, which we are not interested in. Here $\hat{D}_i$ is the covariant derivative with respect to the boundary metric $\gamma_{ij}$.
As expected, the diffeomorphism invariance demands the conservation of the bulk energy-momentum tensor, and the momentum flow condition \eqref{momentumflow} for the plane boundary in the flat space-time.

The invariance under the Weyl transformation $\delta g_{\mu\nu} = 2\sigma g_{\mu\nu}$, which induces $\delta \gamma_{ij} = 2\sigma \gamma_{ij}$, gives a further constraint
\begin{align}
g^{\mu\nu}  T_{\mu\nu}  = \gamma^{ij}  \tau_{ij} = 0 \ . \label{bW}
\end{align}
A simple analogue of Zumino's argument tells that when the boundary condition is Weyl invariant, the restriction to the conformally flat space-time (e.g. above mentioned plane in the Minkowski space-time) will automatically give us the conformal invariant boundary conditions as is obvious from comparing \eqref{traceless} and \eqref{bW}.

If we restrict ourselves to the constant Weyl transformation, the partial integration is possible, so the only constraint is that the boundary energy-momentum tensor is traceless up to a total divergence. 
\begin{align}
\gamma^{ij} \tau_{ij} = \hat{D}^j j_j \ .
\end{align}
This is the origin of the boundary virial current. We stress that the condition for the boundary scale invariance is weaker than the boundary conformal invariance.

\section{Is scale invariant boundary conformal?}
In the last section, we have seen that scale invariant boundary conditions for conformal field theories do not necessarily lead to conformal boundary conditions without further assumptions. As in the bulk, the non-trivial presence of the boundary virial current allows scale invariance without conformal invariance. Indeed, it is easy to construct such a model if we consider non-unitary quantum field theories.

For instance, generic massless vector field theories without gauge invariance (thus without unitarity) are scale invariant but not conformal invariant in any dimension \cite{Riva:2005gd}
\begin{align}
 S= \int d^dx \left(-\frac{1}{4}(\partial_\mu v_\nu - \partial_\nu v_\mu)^2 + \frac{\alpha}{2} (\partial^\mu v_\mu)^2 \right)\ .
\end{align}
with 
\begin{align}
T^{\mu}_{\ \mu} = \left(\frac{d}{2}-2\right)(\partial_\mu v_\nu \partial^\mu v^\nu - \partial_\mu v_\nu \partial^\nu v^\mu) + \alpha\left((2-d)v_\mu \partial^\nu \partial^\mu v_\nu - \frac{d}{2}(\partial^\mu v_\mu)^2 \right) \ .
\end{align}
This can be improved to be traceless only when $\alpha = \frac{d-4}{d}$ (see e.g. \cite{ElShowk:2011gz}). 
Therefore, we can always put such a theory at the boundary and allow a scale invariant coupling to the bulk, leading to scale invariant but not conformal boundary conditions. In the Euclidean signature, this model is regarded as a theory of elasticity, where $v_\mu$ is the displacement vector, so the appearance at the boundary of Euclidean field theories is not at all unnatural.

In this section, we discuss how we may be able to argue that scale invariant boundary conditions lead to conformal boundary conditions with the help of unitarity, causality and some further technical assumptions. 

\subsection{Proof and counterexample in $\mathrm{CFT}_{2}$}
In $(1+1)$ dimensional conformal field theories, we can offer both a proof and a  counterexample that scale invariant boundary conditions will lead to boundary conformal invariance, depending on the assumptions we make. We begin with the proof by carefully noting some crucial steps during the proof, which may need a further justification without assumptions and eventually leading to a counterexample.

``Theorem": Boundary scale invariance in unitary conformal field theories in $(1+1)$ dimension is equivalent to boundary conformal invariance  when (1) the theory has a discrete spectrum, (2) boundary energy-momentum tensor takes a canonical scaling dimension and  (3) the analogue of the Reeh-Schlieder theorem \cite{RS} (i.e. cyclicity of the vacuum) at the boundary is true.

The assumptions (1) and (2) imply that the scaling dimensions of the operators appearing in the theory are diagonalizable. We should note that the third assumption cannot be derived from the other assumptions unlike in higher dimensional quantum field theories where it is possible,  hence known as a theorem. We would like to come back to this point later.

As discussed in section 2, the scale invariance in $(1+1)$ dimensional conformal field theories with boundary (at $r=0)$ implies the existence of the bulk traceless-symmetric-conserved energy-momentum tensor $T_{\mu\nu}$ with the boundary condition $T_{t r}(r=0,t) = \partial^t \tau_{tt}(t)$, where the boundary energy-momentum tensor $\tau_{tt}$ must be the divergence of a virial current $\tau_{tt} = \partial^t j_t$. If $j_t$ can be expressed as $j_t = \partial_t l$, then it is conformal invariant.

A simple dimensional analysis (from the second assumption) tells that $\tau_{tt}$ has scaling dimension one, so  $j_t$ must have scaling dimension zero. If we consider the boundary two-point functions $\langle j_t(t) j_t(s) \rangle $, the scale invariance and translational invariance demands $\langle j_t(t) j_t(s) \rangle = \mathrm{const} $ up to contact terms which are irrelevant for our analysis. The derivatives with respect to $t$ and $s$ yield the vanishing of the two-point function of the divergence of the virial current:
\begin{align}
 \langle \partial^t j_t(t) \partial^t j_t(0) \rangle = 0  \ .
\end{align} 
In higher dimensions, this would imply $\partial^t j_t = 0$ as an operator identity due to the Reeh-Schlieder theorem. In our setup, we cannot derive the Reeh-Schlieder theorem, in part due to the lack of causality. To go further in our argument, we have to explicitly assume the analogue of the Reeh-Schlieder theorem: if $\mathcal{O}(t) |0 \rangle = 0$ then $\mathcal{O}(t) = 0$. Under this assumption, we have shown $\tau^t_{\ t} = \partial^t j_t = 0$ and the scale invariant boundary conditions become conformal invariant. 

The validity of the assumption of the Reeh-Schlieder theorem is somewhat obscure. From the boundary viewpoint alone, it is just a system of quantum mechanics, and there is no simple notion of causality, which makes the proof possible in higher dimension. On the other hand, from the bulk viewpoint, there may be some notions of causality. The same assumption was implicitly made in the so-called ``Hofman-Strominger theorem" that claims that chiral scale invariance in $(1+1)$ dimensional field theories without Lorentz invariance leads to chiral conformal invariance \cite{Hofman:2011zj}. We cannot prove the analogue of the Reeh-Schlieder theorem from their explicit assumptions, so we should have regarded it as an extra implicit assumption, which may not be justifiable \cite{Nakayama:2011fe}. In our case, the condition is obviously necessary because when the boundary is conformal invariant, the unitarity with the state-operator correspondence demands that the Reeh-Schlieder theorem must hold.

One way to justify the assumption would be to imagine that the boundary operators are all given by the limit of certain bulk fields. Then, in the bulk, the extension of the boundary operators must satisfy the microscopic causality, which is the basis of the Reeh-Schlieder theorem, and for its consistency, the microscopic causality must hold for their restriction to the boundary. The argument is reasonable, in particular because it seems that the bulk causality does not seem to be guaranteed otherwise, but we emphasize that this is still an extra assumption.

The above argument does not rely on the existence of $T_{rt}$ at all, so it is valid for scale invariant quantum mechanics, and the above result tells that the  conformal Hamiltonian, which is identified with $ \tau_{tt}$, is always identically zero. This is indeed true for free fermions with the Lagrangian $L =\bar{\psi} \dot{\psi}$. The above argument further tells that the scale invariant boundary condition implies that the momentum flow must vanish at the boundary $T_{rt} (r=0) = 0$. This is nothing but Cardy's boundary condition, and  it eventually allows the conservation of one copy of the Virasoro symmetry.\footnote{We will not dwell on it, but once $T_{rt} = 0$, we can always construct the infinitely many conserved charges according to the standard boundary conformal field theory construction.}

Now let us present a counterexample. The most famous non-trivial conformal quantum mechanics (see e.g. \cite{de Alfaro:1976je}\cite{Chamon:2011xk}) would be the theory with the action
\begin{align}
S = \int dt \left( \dot{\phi}^2 - \frac{g}{\phi^2} \right)
\end{align}
where $\phi$ transforms as a conformal primary with conformal weight $\Delta = -\frac{1}{2}$: 
\begin{align}
\delta_D \phi = t\dot{\phi} - \frac{1}{2}\phi \ , \ \ \delta_K \phi = t^2 \dot{\phi} - t \phi \ .
\end{align}
The action is invariant under dilatation $\delta_D$ as well as the special conformal transformation $\delta_K$.
Suppose we have a boundary conformal field theory which has a boundary conformal primary operator $\mathcal{O}(t)$ whose conformal dimension is $\Delta = \frac{1}{2}$. If we couple these two systems with the scale invariant interaction $\delta S = \epsilon \int dt \dot{\phi} \mathcal{O}$, the conformal invariance is broken due to the fact that $\dot{\phi}$ does not transform nicely under the boundary conformal transformation. Therefore, the coupled system keeps boundary scale invariance but does not acquire conformal invariance. 

In order to reconcile it with the ``theorem", we have to reconsider the assumptions made there. The starting point of the conformal quantum mechanics with the Lagrangian $L = \dot{\phi}^2 - \frac{g}{\phi^2}$ does possess the non-zero Hamiltonian, and the analogue of the Reeh-Schlieder theorem does not hold from the beginning. In addition, the discreteness of the spectrum is questionable since $(\phi \dot{\phi})^n$ are all dimension zero operators. These violations allowed non-zero boundary energy-momentum tensor, which cannot be removed by the improvement. 

We recall that the above ``theorem" also provided a proof that scale invariant quantum mechanics must be conformal invariant under the same assumption (1), (2) and (3). Again, we have a simple ``counterexample" by violating the assumptions. The model is given by the two copies of the conformal quantum mechanics $L = \sum_{i=1}^2 \dot{\phi_i}^2 - \frac{g}{\phi^2_i}$ coupled through the scale invariant interaction $\int dt \epsilon \dot{\phi}_1 \phi_2^{-1}$. We can see that the theory has a non-trivial virial current $j_t = \epsilon \frac{\phi_1}{2\phi_2}$, which cannot be removed by the improvement.

This construction also gives a counterexample of ``Hofman-Strominger theorem". Take the above example of scale invariant but non-conformal quantum mechanics and  let every fields depend on extra coordinate $x$. The theory is obviously translational invariant in $x$ direction, and the scale invariance is intact by assigning zero weight for $x$ coordinate, as is compatible with their chiral scale algebra. The theory is not chiral conformal invariant, and this is a counterexample of their claim. We can use whichever scale invariant but non-conformal quantum mechanics here. As we have discussed, the problem is that the Reeh-Schlieder theorem cannot be proved within their assumptions, and our model indeed violates the cyclicity of the vacuum.

\subsection{Necessary and sufficient conditions in $\mathrm{CFT}_3$}
We move on to the higher dimensional cases. In $(1+2)$ dimensional conformal field theories with boundaries, we can make more precise the condition for the enhancement of conformal invariance from scale invariance. 

The idea is to study the structure of the boundary two-point functions. We introduce the complex notation for the boundary coordinate $z = x + iy$ and $\bar{z} = x-iy$ up on the Wick rotation $t \to iy$. The boundary energy-momentum tensor has only two independent components: $\tau = \tau_{zz} = (\tau_{\bar{z}\bar{z}})^*$ and $\theta = \tau^{\mu}_{\ \mu}$. In addition, we need to introduce a particular component of the bulk energy-momentum tensor $T=T_{rz} = (T_{r\bar{z}})^*$ evaluated at the boundary $r=0$.

With the same assumption that the spectrum of the dilatation operator is discrete and diagonalizable, the structures of the boundary two-point functions are largely determined from the scaling analysis. In the momentum space by dropping the delta functions for the momentum conservation, we have
\begin{align}
\langle \tau(k) \tau(-k) \rangle = c \frac{k^3}{\bar{k}} \ , \ \ \langle \tau(k) \bar{\tau}(-k) \rangle = f |k|^2 \log |k|^2 \cr
\langle \tau(k) \theta(-k) \rangle = e k^2 \log |k|^2  \ , \ \ \langle \theta(k) \theta(-k)  \rangle = h |k|^2\log|k|^2 \cr
\langle \tau(k) T(-k) \rangle = x k^3 \log |k|^2  \ , \ \ \langle \theta(k) T(-k)  \rangle = y k|k|^2\log|k|^2 \cr 
\langle  T(k) T(-k)  \rangle = z k^2 |k|^2 \log |k|^2  \ , \ \ \langle T(k) \bar{T}(-k)  \rangle = w |k|^4\log|k|^2 \cr
\langle \bar{\tau}(k) T(-k)\rangle = l \bar{k}|k|^2  \log |k|^2 
\end{align}
Here, we have neglected the contact terms. Unitarity demands $c \ge 0$, $ z\ge 0$ and $h \ge 0$.
Note the absence of the term $ \frac{k^3}{\bar{k}} \log |k|^2$ in $\langle \tau(k) \tau(-k) \rangle$  due to our assumption of canonical scaling dimensions \cite{Polchinski:1987dy}\cite{Dorigoni:2009ra}.
 Each two-point functions only have a unique term in the momentum space.
 We impose the ``conservation" equation $\bar{\partial} \tau + 4\partial \theta  = T$ at the boundary. It determines the relation among coefficients as
\begin{align}
 4e = x \ , \ \ e + 4h = y \ , \ \  f + 4e^* = l \ , \ 8e + 16h = z \ , \ \
  f + 4 e + 4e^* + 16h = w \ .
\end{align}

This is all the information encoded in the symmetry of the scale invariance and  the momentum flow condition. Unlike in bulk $(1+1)$ dimensional scale invariant field theories, we cannot claim the conformal invariance (i.e. $e=0$), but we can still infer a non-trivial necessary and sufficient condition for the boundary conformal invariance.
First of all, we note that Cardy's condition $T_{rz} = T_{r\bar{z}} = 0$ is a sufficient condition for the conformal invariance. This is because when $T=0$, the conservation equation tells that $ e = \langle \theta(k) \theta(-k) \rangle = 0$ up to contact terms. From the unitarity, causality and the discreteness of the spectrum, the Reeh-Schlieder theorem holds in boundary $(1+1)$ dimension, so we can conclude $\theta = 0$ as an operator identity and the boundary theory is conformal invariant.

Is Cardy's condition necessary for conformal invariance? Assume $e = h = y = 0$. The boundary conservation equation demands $x = z =0$. In particular $\langle T(k) T(-k) \rangle = 0$,  so the Reeh-Schlieder theorem dictates $ T_{rz} = {T}_{r\bar{z}} = 0$. Alternatively, we  can see this result by noting that $\tau_{ij}$ is a symmetric traceless tensor with conformal dimension two, and the unitarity of the boundary conformal algebra demands such an operator must be conserved \cite{Mack:1975je}\cite{Minwalla:1997ka} and $T_{ri}= 0$. 

We therefore conclude that in $(1+2)$ dimension, Cardy's condition $T_{ri}=0$ is a necessary and sufficient condition for the boundary conformal invariance. We do not yet know how this can be derived from the scale invariance alone, which will be the central discussions from the perturbative approach in section 3.3 and the holographic approach in section 4. 

As a corollary, as long as the boundary condition continues to satisfy Cardy's condition along the boundary renormalization group flow, one can always construct a $g$-function that decreases along the boundary renormalization group flow. It can be simply constructed by the same two-point functions considered by Zamolodchikov \cite{Zamolodchikov:1986gt} because only the conservation equation is relevant in the proof there, and the same argument applies here.

In higher dimensions $d>1+2$, Cardy's boundary condition is apparently not sufficient within the two-point function analysis suitably extended. This can be clearly seen from the fact that we may be able to put an independent scale invariant but non-conformal field theory at the boundary with no interaction to the bulk {\it assuming its existence}, which cannot be forbidden from the two-point function analysis in this section. Unfortunately, we do not know any good example of such (with the Noether assumption and unitarity), so it is not conclusive. With a slight relaxation of the Noether assumption, free $U(1)$ Maxwell theory can be regarded as such when $d >4$ (see e.g. \cite{ElShowk:2011gz}).

However, under the assumption of canonical scaling dimension of the boundary energy-momentum tensor and the unitarity, we can show Cardy's boundary condition $T_{r\mu}=0$ is actually necessary. The reason is that conformal invariance requires that $\tau_{\mu\nu}$ is a symmetric traceless tensor whose conformal dimension is $d-1$. The unitarity of the boundary conformal algebra on the other hand demands that such an operator must be conserved. Then Cardy's condition  $T_{r\mu} = 0$ follows.

\subsection{Perturbative fixed points and $g$-function}
In order to understand the relation between scale invariance and conformal invariance, the generalization of Zamolodchikov's $c$-theorem has played a crucial role in the bulk. The higher dimensional analogue of the $c$-theorem relates the higher point correlation functions of the energy-momentum tensor or conformal anomaly to the monotonically decreasing function along the renormalization group flow \cite{Cardy:1988cwa}\cite{Komargodski:2011vj}. The perturbative  evaluation of the rate of the change of the generalized $c$-function leads to the conclusion that the beta function must vanish at the perturbative renormalization group flow fixed point and the theory must be not only scale invariant but also conformal invariant. In this subsection, we would like to utilize the same idea to argue that at the perturbative fixed points of the boundary theory, the boundary condition is not only scale invariant but also conformal invariant.

In boundary field theories, the analogue of the $c$-theorem is proposed as $g$-theorem, claiming that there exists a $g$-function that monotonically decreases along the renormalization group flow \cite{Affleck:1991tk}\cite{Friedan:2003yc}.\footnote{As in the bulk case \cite{Zamolodchikov:1986gt}\cite{Polchinski:1987dy}, the proof of the $g$-theorem in $(1+1)$ dimension in \cite{Friedan:2003yc} should imply the enhancement of boundary conformal invariance in $(1+1)$ dimension with a minor extension of their argument. Our proof in section 2 had a slightly different taste (e.g. the topology of the boundary is different, and the integration over the space-time is not needed). The zero temperature case was also discussed in \cite{Friedan:2005dj}, which is closer to our discussion.
The author would like to thank A.~Konechny for discussions and clarification.} Unfortunately, the nature of $g$-function in higher dimensions is not well-understood. Our approach is based on the first order conformal boundary perturbation theory, and use the proposal made in \cite{Nozaki:2012qd} that relates the $g$-function and the conformal anomaly of the partition function of the $d$-dimensional round ball $B_d$ with boundaries $S_{d-1}$ when $d$ is an odd integer. 

One disadvantage of this approach is that we have to consider the curved background rather than the Minkowski space-time with the plane boundary we started with in section 2. These are related by conformal mapping, so it is irrelevant for the study of the conformal invariance, but we have to remind ourselves that with the scale invariance alone, there is no direct connection between two setups. Nevertheless, as we will see that the $g$-function only depends on the local properties of the renormalization group structure at the order we are interested in, so we hope that the subtlety does not affect our final conclusion. In section 4, we pursue the holographic approach where the Poincar\'e symmetry is kept intact and there should be a similar construction in the field theory side, but we leave it for a future study.

Let $r_B$ be the radius of the ball $B_d$ and impose a suitable boundary conditions at the boundary $S_{d-1}$. The proposed $g$-function \cite{Nozaki:2012qd} is given by the renormalized partition function $\log Z_{B}$. The claim of the proposed $g$-theorem is if we define $g$ by
\begin{align}
g(r_B) = 3(-1)^{\frac{d+1}{2}} \frac{d\log Z_{B}}{d\log r_B} 
\end{align}
then $g$ is monotonically decreasing
\begin{align}
\frac{d g(r_B)}{d \log r_B} \le 0 \ 
\end{align}
when $d$ is odd. When $d$ is even, one may take the finite part of the ball partition function as the $g$-function.

We assume that the theory is given by the perturbation of the boundary conformal field theory with some relevant deformations
\begin{align}
S = S_{\mathrm{BCFT}} + g^I \int_{S_{d-1}} dV \mathcal{O}_I(x) \ ,
\end{align}
where $\mathcal{O}_I$ are boundary conformal primary operators. For the following argument, it is important that $\mathcal{O}_I$ are primary operators. In particular, there is no relevant deformation from the descendant of vector operators such as $\partial^i  j_i$ : their scaling dimension must be greater than $d-1$ from unitarity (and after all it must vanish up on integration).

We will define the beta-functions for the renormalized coupling $g^I$ by
\begin{align}
\beta^I = \frac{dg^I(\mu)}{d\log \mu} \ ,
\end{align}
which can be computed by the conformal perturbation theory. 
\begin{align}
\beta^I = -(d-1-\Delta) g^I + \frac{\pi^{\frac{d-1}{2}}}{\Gamma(\frac{d-1}{2})}C^{I}_{\ JK} g^J g^K + \mathcal{O}(g^3) \ ,
\end{align}
where $C^{IJK}$ is the OPE coefficient (raised and lowered by the boundary Zamolodchikov metric $G_{IJ}$). 
Since the vector descendant operators cannot appear in the relevant part of the OPE from the unitarity as we mentioned, there is no distinction between beta function and $B$ function (which appears in the total trace of the boundary energy-momentum tensor as $\tau^i_{\ i} = B_I \mathcal{O}^I = \beta_I \mathcal{O}^I - \partial^i J_i$) at the lowest order in conformal perturbation theory. We will say more about the $B$ function later in the last two paragraphs of this section.

With the use of the beta functions, the first order change of the partition function was computed as
\begin{align}
\frac{d}{d\log r_B} \left((-1)^{\frac{d+1}{2}} \frac{d\log Z_B}{d\log r_B}\right) = -\frac{2\pi^{d-\frac{1}{2}}}{\Gamma(\frac{d+1}{2})\Gamma(\frac{d}{2})2^{d-2}} G_{IJ} \beta^{I} \beta^{J} + O(g^4)  \ ,
\end{align}
where $G_{IJ}$ is again the boundary Zamolodchikov metric evaluated at the unperturbed conformal fixed point, which is manifestly positive definite at the conformal fixed point from unitarity. This shows that $g$-theorem holds at the first non-trivial order in perturbation theory \cite{Nozaki:2012qd}. 
When $d$ is even, one can compute the finite part of the renormalized partition function in a similar way, and the boundary perturbation leads to the similar conclusion that can be found in \cite{Klebanov:2011gs} because the bulk does not play any role at the order they studied.

At the scale invariant fixed point, the renormalized partition function must be constant up to the logarithmic term with a fixed coefficient when $d$ is odd:
\begin{align}
\log Z_B = \frac{g_*}{3} \log\frac{r_B}{\epsilon} \ .
\end{align}
This means that at the scale invariant fixed point $\beta_I = 0$. On the other hand, we may be able to expand the trace of the boundary energy-momentum tensor with respect to $\mathcal{O}^I$ by $\tau^{i}_{\ i} = \beta^I O_I$ since the trace of the boundary energy-momentum tensor is the response to the Weyl transformation of the boundary.\footnote{We again emphasize at the lowest order in conformal perturbation theory with the prescription by \cite{Nozaki:2012qd}, there is no distinction between beta functions and $B$ functions. However at the higher order in perturbation theory, they may deviate from each other. See the following arguments.}  The argument here is equivalent to the non-perturbative argument in $(1+1)$ dimension given in \cite{Friedan:2003yc}.
Thus as long as $G_{IJ}$ is non-degenerate, the perturbative scale invariant fixed point of the boundary theory is a boundary conformal field theory. 

In the above discussions, we have assumed that the deformation is induced by relevant operators. 
It should be possible to derive a similar $g$-function for marginally relevant deformations (or marginally irrelevant deformations by reversing the renormalization group flow). In this case, we have to be more careful about the mixing of the marginally relevant deformations with the descendant of the vector primary operators whose dimension is $d-1$. This leads to the operator identity like $(S \cdot \beta)_I\mathcal{O}^I = \partial^i J^{(S)}_i$ and gives rise to the ambiguity of the beta functions \cite{Jack:1990eb}\cite{Osborn:1991gm}. In particular, in this situation, the trace of the boundary energy-momentum tensor should be expanded with the gauge invariant $B$ function, which is defined by $B_I \mathcal{O}^I = \beta_I \mathcal{O}^I - \partial^i J_i$ so that the above mentioned ambiguity cancels: $\beta_I \to \beta_I + (S\cdot \beta)_I$ and $J_i \to J_i + J^{(S)}_i$ will be understood as the gauge invariance of the background field (which will be more manifest in holographic approach later).\footnote{Probably, one quick way to convince ourselves of the ambiguity is to consider the open string perturbation theory, where $B$ function for the gauge field vertex operator (at one-loop) is $B_{A_\mu} =\partial^\nu F_{\mu\nu} $, but for finiteness of the perturbation, there is nothing wrong with adding extra ``gauge transformation" so that $\beta_{A_\mu} =\partial^\nu F_{\mu\nu} + \partial_\mu\Lambda$. We know that Weyl (conformal) invariance dictates $B=0$ rather than $\beta =0$ here.} 
One can compute the beta functions by demanding the finiteness of the renormalized coupling constants. The computation of the partial virial current $J_i$ is more difficult because the flat space scale transformation cannot detect it, and we have to consider the Weyl transformation, or position dependent coupling constant. In the perturbative bulk quantum field theories, the technique can be found in \cite{Jack:1990eb}\cite{Fortin:2012hn}.
In our application, it is crucial that we have to show that $B$ functions rather than beta functions vanish at the scale invariant fixed point to argue the boundary conformal invariance.

It is not entirely obvious if the ball partition function captures this subtlety because we had to keep track of all the total derivative terms in the effective action to see the effect, but  in the conformal perturbation theory at the lowest order discussed in this subsection, the divergence of the vector operator with dimension $d-1$ is necessarily zero at the conformal fixed point from unitarity, so there is no distinction between beta functions and $B$ functions at this order. We will discuss in section 4 that the holographic $g$-function encodes this subtlety and indeed predicts that $B$ function must vanish at the scale invariant fixed point, and the boundary theory is conformal invariant as long as the analogue of the Zamolodchikov metric is positive definite.

\section{Holographic approach}
In order to gain more intuition on the boundary conditions for conformal field theories, and in particular to understand possible enhancement of boundary conformal invariance, we will study the holographic approach, which is complementary to the perturbative approach discussed in section 3.4. Although the applicability of the holographic technique may be limited, in the bulk case, it has been shown that the general constraint on the renormalization group flow is beautifully realized and suggests a proof of the enhancement \cite{Nakayama:2009qu}\cite{Nakayama:2009fe}\cite{Nakayama:2010wx}\cite{Nakayama:2010zz}. We will expect that the similar argument applies in the boundary theories.

\subsection{Basic setup for holographic BCFT}
Let us first begin with the construction of the holographic boundary conformal field theory proposed in \cite{Takayanagi:2011zk}\cite{Fujita:2011fp}\cite{Nozaki:2012qd}. For this purpose, in the gravity sector, we impose the Dirichlet boundary condition at the AdS boundary as usual in the AdS/CFT correspondence while we impose the Neumann boundary condition at the new boundary where the conformal boundary theory is holographically realized.\footnote{Boundary conformal field theories in holography were also studied in \cite{Karch:2000ct}\cite{Karch:2000gx}. More recent references include \cite{Alishahiha:2011rg}\cite{Kwon:2012tp}\cite{Fujita:2012fp}. More top-down oriented approaches can be found in \cite{Chiodaroli:2011fn}\cite{Chiodaroli:2012vc}. AdS boundary condition was also studied in\cite{Andrade:2011nh}.}

We will follow the notation of \cite{Nozaki:2012qd}. We take the total space $N$ to be a locally $AdS_{d+1}$ space-time with the $AdS_{d+1}$ boundary $M$ (Dirichlet boundary condition), where the bulk conformal field theory  ``lives", and the other bulk boundary $Q$ (Neumann boundary condition) that divides the original $AdS_{d+1}$ space-time. We also call $P$ the intersection of two boundaries $M$ and $Q$, where the boundary conformal field theory ``lives". To preserve the boundary conformal invariance, we assume that the total space is foliated by $AdS_{d}$, and therefore the boundary $Q$ is locally $AdS_{d}$ space-time.

The gravitational part of the holographic action is given by the usual Einstein-Hilbert term with cosmological constant and the Gibbons-Hawking term
\begin{align}
 S = \frac{1}{2}\int d^{d+1}x \sqrt{g}  \left(R-2\Lambda\right) + \int_Q \sqrt{h} K + \int_{M} \sqrt{h}K \ ,
\end{align}
where $K$ is the extrinsic curvature scalar for each boundaries $Q$ and $M$.

As is discussed in \cite{Nozaki:2012qd}, we have to further introduce codimension two boundary terms as well as various counter-terms to remove divergences in holographic renormalization group. In particular, we have to add the codimension two boundary term:
\begin{align}
S_P = \int_P \sqrt{\Sigma} (2\theta -\pi) \ , \label{tbt}
\end{align}
where $2\theta$ is the angle between $Q$ and $M$ at $P$, and $\Sigma$ is the induced metric on $P$. We have neglected further curvature terms on $P$ which is only relevant when we consider the extrinsic curvature of the field theory boundary.
 We will furthermore add non-trivial matters on $Q$ whose energy-momentum tensor is denoted by $T_{ab}^{(Q)}$ so that the Neumann boundary condition gives the boundary Einstein equation (we will use $x^a$ for the coordinate on $Q$)
\begin{align}
K_{ab} - \gamma_{ab} K = T_{ab}^{(Q)} \ . \label{bE}
\end{align}

If we would like to construct a hologrpahic dual of a flat space-time conformal field theory with the plane conformal boundary, the geometry is uniquely specified by the metric
\begin{align}
ds^2 = d\eta^2 + \cosh^2(\eta/L)\frac{dz^2 + \eta_{ij}dx^i dx^j}{z^2}\ . \label{bmetric}
\end{align}
where the boundary $Q$ is located at $\eta = \eta^*$ and the $AdS_{d+1}$ radius $L$ is given by $L = \sqrt{\frac{\Lambda}{d(d-1)}}$. The position of the boundary is specified by the boundary cosmological constant through the boundary Einstein equation \eqref{bE} and gives
\begin{align}
\frac{\eta^*}{L} = \mathrm{Arctanh}(\frac{TL}{d-1}) \ , \label{tension}
\end{align}
where we have assumed $T = \gamma^{ab}T_{ab}^{(Q)}$ takes a constant value provided by the boundary cosmological constant or tension of the boundary. Physically, the tension of the boundary determines the number of the degrees of freedom encoded in the holographic boundary at $P$.

To define the holographic energy-momentum tensor, we allow a small fluctuation of the geometry around the vacuum solution \eqref{bmetric}.
Suppose we have the following form of the near AdS boundary behavior given by the Fefferman-Graham expansion
\begin{align}
ds^2 = \frac{L^2}{4\rho^2} d\rho^2 + \frac{1}{\rho} g_{\mu\nu}(x,\rho) dx^\mu dx^\nu
\end{align}
The holographic bulk energy-momentum tensor is defined as 
\begin{align}
T^{(\mathrm{hol})}_{\mu\nu}  = \lim_{\rho \to 0} \left({\rho^{1-\frac{d}{2}}}(K_{\mu\nu} - \gamma_{\mu\nu} K) + (\text{counter terms}) \right) \ , \label{bulkem}
\end{align}
where $K_{\mu\nu}$ is the extrinsic curvature on $M$ with the boundary metric $\gamma_{\mu\nu}$ which we fix.

Similarly the holographic boundary energy-momentum tensor is defined by
\begin{align}
\tau_{ij}^{(\mathrm{hol})} = \lim_{\rho \to 0} \left(\rho^{\frac{3}{2}-\frac{d}{2}}(2\theta -\pi) \Sigma_{ij} + \text{counter terms} \right) \ . \label{boundaryem}
\end{align}
Generically, these expressions \eqref{bulkem} \eqref{boundaryem} are divergent, and we need to regularize the expression by introducing the counterterms. Only the finite or logarithmic terms are relevant quantities universally comparable with the quantum field theory results. If we are interested in the non-universal terms, the counterterms in \eqref{bulkem} and \eqref{boundaryem} must be related  in order to satisfy the conservation equation.

It was observed in \cite{Nozaki:2012qd} that the Fefferman-Graham expansion needs a suitable modification, when the boundary does not preserve the boundary Poincar\'e symmetry in higher dimension $d>2$. Since we are only interested in the Poincar\'e invariant boundary condition in tangential direction, this is not important for us.

By studying the diffeomorphism invariance of the gravity action, one can derive the conservation of the boundary energy-momentum tensor up to extrinsic curvature terms we will neglect
\begin{align}
T_{r i}^{(\mathrm{hol})} = \partial^j \tau_{ji}^{(\mathrm{hol})} \ .
\end{align}
The extrinsic curvature corrections to the conservation equation should have been obtained by adding further terms in the boundary terms \eqref{tbt} on $P$, which we will not pursue further, and we have already discarded in \eqref{conserve}. We are only interested in the plane boundary condition, so the correction, although important when the boundary has an extrinsic curvature, is irrelevant for our discussions.

Since the boundary condition preserves the $AdS_{d}$ isometry, the vacuum solution \eqref{bmetric} preserves the boundary conformal invariance. The unitarity argument in the previous section suggests that Cardy's boundary condition $T_{ri} = 0$ must be satisfied. We can easily see that with the plane boundary condition $g_{\mu\nu} = \eta_{\mu\nu}$ and $\eta(z) = \eta^*$, we obtain
\begin{align}
\langle T_{r i}^{(\mathrm{hol})} \rangle = 0 \ 
\end{align} 
so Cardy's condition is satisfied as an expectation value.

\subsection{Boundary scale vs conformal invariance}
Now we would like to consider a generalization of the holographic boundary conformal field theory construction. Can we realize the scale invariant but non-conformal boundary conditions for a conformal field theory as a generalization of the holographic approach? For the discussion, we assume that the bulk is locally $AdS_{d+1}$ space-time because we assume the bulk conformal invariance.\footnote{Again we stress there is no meaning to discuss the conformal invariance of the boundary when the bulk is not conformal invariant.} 
We also assume that the boundary preserved scaling invariance is realized by the isometry $ (z \to \lambda z, x_i \to \lambda x_i)$ while the Poincar\'e invariance acts on $x_i$ in the usual manner.

Let $\eta = \eta^*$ be the boundary $Q$ of the bulk space-time $M$, where the boundary conformal field theory is holographically realized and we put the Neumann boundary condition at $\eta = \eta^*$ as in section 4.1.  The most generic metric compatible with the above assumption is
\begin{align}
ds^2 = a(\eta) d\eta^2 + \frac{f(\eta) dz^2 + g(\eta) \eta_{ij} dx^i dx^j}{z^2} + h(\eta) \frac{dz d\eta}{z}\ .
\end{align}
One can always (locally) assume $a(\eta) = 1$ and $h(\eta) = 0$ by changing the coordinate $z \to \alpha(\eta) z$ and $\eta \to \beta(\eta)$ without spoiling the assumed isometry. In addition, from the assumption that the bulk geometry must be locally $AdS_{d+1}$, we can conclude that $f(\eta)/g(\eta) = \mathrm{const}$, and the constant can be chosen to be $1$ by rescaling $x^i$. The resulting geometry is nothing but the one studied in the last subsection \eqref{bmetric}. The condition of the local $AdS_{d+1}$ space-time further determines $f(\eta) = \cosh^2(\eta/L)$. The value of $\eta^*$ is given by the boundary cosmological constant \eqref{tension}. In this way, we have obtained the space-time that is foliated by $AdS_d$ in which the boundary conformal invariance is manifestly realized even though we did not impose the invariance under the boundary special conformal transformation.

The field theoretic interpretation is rather clear. The minimal assumption of the scale invariant boundary condition requires the existence of the non-trivial boundary virial current $\tau^{i}_{\ i} = \partial^i j_i$ for the trace of the boundary energy-momentum tensor. However, in the geometric construction here with the full general covariance, there is no candidate for the gravity dual of the virial current $j_i$ within the pure gravity sector.\footnote{The full diffeomorphism is crucial here. Once we abandon the full diffeomorphism and only keep the foliation preserving diffeomorphism in holographic $z$ direction, we will have a non-trivial candidate for the dual of the boundary virial current, and the boundary conformal invariance will be lost.} Thus, the geometric realization of the boundary scale invariant condition must preserve the boundary conformal invariance or $AdS_d$ isometry. The same argument appeared in the bulk enhancement of conformal invariance from scale invariance within the pure gravity sector. 

To obtain the scale invariant but non-conformal gravity dual, let us consider introducing the boundary fields localized at $\eta = \eta^*$. These localized field can break the $AdS_d$ isometry of the boundary. For instance, suppose we have a vector field whose configuration is $A = \frac{\alpha dz}{z}$ on Q. Such a field configuration is scale invariant but not conformal invariant under the special conformal $AdS_d$ isometry
\begin{align}
\delta z &= 2(\rho^j x_j) x \cr
\delta x^i &= 2(\rho^j x_j)x^i - (z^2 + x^j x_j)\rho^i \ . \label{spec}
\end{align}

We would like to argue that the boundary vector field condensation $A = \frac{\alpha dz}{z}$ on Q corresponds to the existence of the  boundary virial current in the trace of the boundary energy-momentum tensor. For this purpose, let us reconsider the boundary beta functions and the possibility to rewrite it as a divergence of the boundary virial current due to operator identities:
\begin{align}
\tau^{i}_{\ i} = B^I O_I = \partial^i j_i \label{tbe}
\end{align}
To realize the operator identity $B^I O_I = \partial^i j_i$ in holography, we introduce the boundary scalar fields $\Phi^I$ (dual to $O_I$) and the boundary gauge fields $A_M$ (dual to $j_i$) living on $Q$ (the bulk version of the argument can be found in appendix of \cite{Nakayama:2012sn}).  We assume $\Phi^I$ transforms non-trivially under the gauge transformation of $A_M$ e.g. in the Abelian case $A \to A + d\Lambda$, $\Phi^I \to e^{-iq_I \Lambda} \Phi^I$. 

Suppose at the vacuum, the gauge symmetry $A \to d\Lambda$ is spontaneously broken by the condensation of $\Phi^I$. In addition,  we assume that the vacuum configuration is scale invariant as well as Poincar\'e invariant. If we choose a particular gauge, the configuration will look like $\Phi^I = a^I z^{i q_I \alpha}$ with $A =0$. This corresponds to the non-zero $B^I$ function in  \eqref{tbe} because it means that the coupling constant $g^I$ runs under the holographic renormalization group flow. The resulting holographic boundary renormalization group flow looks like cyclic (see also \cite{Nakayama:2011zw}).

On the other hand, we can choose a different gauge $\Phi^I = a^I$ with $A = \alpha \frac{dz}{z}$. This gives a holographic interpretation of the right hand side of the equation \eqref{tbe}. Of course, we can also choose a mixed gauge so that we recover the ambiguities in the beta functions discussed in \cite{Jack:1990eb}\cite{Osborn:1991gm} by transforming it to the divergence of the (partial) virial current. 
As we have advertised, the ambiguity of the beta functions with the use of the operator identity is nothing but the gauge transformation in the gravity dual.

We have a couple of comments about the construction. Suppose we have a gravity dual of a boundary conformal field theory. It may contain a spontaneously broken gauge symmetry with $\Phi^I = a^I$ and $A=0$. The theory is manifestly conformal invariant. However, if we perform the gauge transformation, we have $\Phi^I = a^I z^{-iq_I \alpha}$ {\it and} $ A = \alpha \frac{dz}{z}$. The configuration is not invariant under the naive special conformal $AdS_{d}$ transformation \eqref{spec}, but it should be gauge equivalent. This is the dual gravitational configuration for the apparently non-conformal but scale invariant field theories studied in \cite{Fortin:2012cq}\cite{Fortin:2012hn}, which turn out to be hiddenly conformal invariant.\footnote{As a consequence, the lack of the conformal invariance in their models are gauge artifact from the holographic perspective.} The precisely same construction applies to the boundary theory here.

In order to compare the holographic argument with the perturbative field theory construction, it is necessary to fix the gauge to define the holographic (boundary) renormalization group flow. The most convenient choice and the one used in the most part of this paper is to set $A=0$ (unitary gauge). In the field theory language, the boundary renormalization group flow is defined by the gauge invariant $B$ functions. 

In such a situation, we know that the Higgs mechanism makes the gauge field massive. Therefore, according to the standard AdS/CFT recipe, the vector operator $j_i$ possess the anomalous dimension as it should be. Otherwise, the unitarity of the boundary conformal algebra demands $\partial^i j_i = 0$ (at least $d>2$). On the other other hand, $\tau^{i}_{\ i}$ must have conformal dimension $(d-1)$ with no anomalous dimension. Of course, this is not a contradiction because only the difference $\partial^i j_{i} - \beta^I O_I$, which turns out to be zero for boundary conformal field theories, must have the conformal dimension $(d-1)$, and we can always declare the operator ``$0$" has dimension $(d-1)$.

In genuine scale invariant but non-conformal invariant situations, the Higgs mechanism gives an interesting suggestion of the violation of unitarity in such models from holography. As in the above, if the Higgs mechanism occurs in unitary  field theories, the gauge field acquires a non-zero mass, thus the (boundary) virial current must have an anomalous dimension. However, it must be equated with the trace of the (boundary) energy-momentum tensor which cannot have anomalous scaling dimension, and unlike in the conformal case, there is no cancellation that saves the discrepancy.
We may avoid the inconsistency by postulating Higgs mechanism without making gauge field massive. This will lead to the violation of  the so-called strict null energy condition as we will see in the next section, and most probably, such a theory is not unitary.\footnote{In the flat space-time, the Proca condition that removes the negative norm states is precisely due to the mass of the vector field, and without gauge invariance, zero-mass limit is singular.}

\subsection{Holographic proof}
As in the holographic proof of the enhancement of bulk conformal invariance from bulk scale invariance, our approach to prove the holographic enhancement of boundary conformal invariance from boundary scale invariance relies on the holographic counterpart of the ``$g$-theorem" \cite{Takayanagi:2011zk}\cite{Fujita:2011fp}.

To understand the holographic $g$-theorem, we consider putting a boundary in $AdS_{d+1}$ space with the coordinate
\begin{align}
ds^2 = d\eta^2 + \cosh^2(\eta/L)\frac{dz^2 + \eta_{ij}dx^i dx^j}{z^2}\ .
\end{align}
at $\eta = \eta(z)$. We know the conformal boundary condition demands that the boundary is located at constant $\eta = \eta^*$ determined by the boundary cosmological constant. We imagine that the boundary renormalization group is realized as a non-trivial profile of $\eta(z)$ induced by a non-trivial matter dynamics on $Q$.

Following the discussions of \cite{Takayanagi:2011zk}\cite{Fujita:2011fp}, we introduce the $AdS_{d+1}$ Poincar\'e coordinate 
\begin{align}
w = \frac{z}{\cosh(\eta/L)} \ , \ \ \xi = z \tanh(\eta/L) \ 
\end{align}
so that (with the rescaling of $x^i$)
\begin{align}
 ds^2 = L^2 \frac{dw^2 + d\xi^2 + \eta_{ij} dx^i dx^j}{w^2} \ .
\end{align}
In the new coordinate we can parametrize the boundary $\xi = \xi (w)$, and the conformal boundary condition is equivalent to $\xi = \sinh(\eta^*/L) w$. 
We now define the holographic $g$-function as\footnote{We use a slightly different definition of the holographic $g$-function than the one presented in \cite{Takayanagi:2011zk}\cite{Fujita:2011fp}.  Their choice corresponded to $\log g(z) = \eta (z)$. Ours and theirs both satisfy the monotonicity and agree at conformal fixed points, so it must be related by the scheme change. Ours corresponds to a holographic scheme \cite{Erdmenger:2001ja}. See also \cite{Kwon:2012tp} for a choice similar to ours.}
\begin{align}
\log g(w) = \mathrm{Arcsinh}\left(\frac{d\xi(w)}{dw} \right) \ .
\end{align}
We recall that $w$ is the natural renormalization group direction for the bulk fields.
Since arcsinh is a monotonic function, the monotonicity of the holographic $g$-function under the renormalization group flow is governed by $\frac{d^2\xi(w)}{dw^2}$.
Applying the boundary Einstein equation with the presence of the matter localized at the boundary $Q$:
\begin{align}
(K_{ab} - \gamma_{ab}K) = T_{ab}^{(Q)} 
\end{align}
 we obtain (in the following $\xi' = \frac{d\xi(w)}{dw}$)
\begin{align}
-\frac{d^2 \xi(w) }{dw^2} = z(1+(\xi')^2)^{3/2} K_{ab} k^a k^b  \ .
\end{align}
where we have introduced a particular null vector $k^a$ 
\begin{align}
k^t = 1 \ , \  \ k^w = \frac{1}{\sqrt{1+(\xi')^2}} \ , \ \ k^{\xi} = \frac{\xi'}{\sqrt{1+(\xi')^2}} \ . \label{nullk}
\end{align}
Note that the null vector is tangent to the boundary.
Suppose we impose the null energy condition on $T_{ab}^{(Q)}$, then
\begin{align}
\frac{d\log g(w)}{d\log w} \sim - T_{ab} k^a k^b \le 0  \ 
\end{align}
up to a manifestly positive function in the right hand side.
Therefore the null energy condition guarantees that $\log g(w)$ is monotonically decreasing along the renormalization group flow, and it provides the holographic proof of the $g$-theorem. Physically, along the renormalization group, the boundary loses the degrees of freedom and the tension of the boundary decreases.

In order to relate the similar discussions in section 4, we consider the situation in which the matter at the boundary $Q$ is explicitly given by the non-linear sigma model whose metric is $G_{IJ}$. The boundary energy-momentum tensor appearing in the holographic $g$-theorem is given by
\begin{align}
T_{ij} k^i k^j = G_{IJ} \partial^{\hat{w}} \Phi^I \partial_{\hat{w}} \Phi^J \ ,
\end{align}
where $\hat{w}$ is the tangential direction at the boundary $Q$ (the spatial direction of \eqref{nullk}). The right hand side is positive definite when the target space has a positive definite metric. It is reasonable to interpret that $\partial^{\hat{w}} \Phi^I$ is the rate of the change of the boundary coupling constant $g^I$ dual to $\Phi^I$ along the renormalization group flow, i.e. beta functions.
Note that in the previous section, we used $z$ as the renormalization group direction, while here we use its tangential projection $\hat{w}$ on $Q$. In the last section, the choice was unique due to the scale invariance, but here the difference comes from the lack of the scale invariance, and we have some ambiguities. This does not affect our argument at all, and near the conformal fixed points, the difference, which is nothing but the scheme dependence, is small. 
 The interpretation leads to the holographic proof of the strong $g$-theorem
\begin{align}
\frac{d\log g(\mu)}{d\log\mu} = -G_{IJ} \beta^I \beta^J
\end{align}
which can be compared with the perturbative result in section 3.3. Here, it seems natural to assume the positivity of the metric because otherwise the unitarity of the holographic dual is at risk.

In order to study the relation between boundary scale invariance and conformal invariance, we recall that we can introduce a relevant operator identity between the $B$ functions and candidate of the virial current by gauging the non-linear sigma model. The holographic $g$-theorem still reads
\begin{align}
\frac{d\log g(w)}{d\log w} \sim - T_{ab} k^a k^b = - G_{IJ} D^{\hat{w}} \Phi^I D_{\hat{w}} \Phi^J 
\end{align}
Now it seems clear that whenever $G_{IJ}$ is positive definite, the would-be scale invariant but non-conformal field configuration gauge equivalent to $A = \alpha \frac{dz}{z}$ with $\Phi^I = \mathrm{const}$ is incompatible with the scale invariant boundary condition. In other words, whenever the $B$ functions are non-zero, the holographic $g$-function is monotonically decreasing along the renormalization group flow, and never reaches the renormalization group fixed points (or trajectories).

The null energy condition alone does not forbid the degenerate metric for the non-linear sigma model, but from the unitarity requirement, it seems mandatory to require the stronger condition that forbids the degenerate metric. One sufficient condition is the so-called strict null energy condition \cite{Nakayama:2010wx}: whenever the field configuration saturates the null energy condition, it must be trivial in the sense that it is invariant under the isometry of the space-time. 

With the help of the strict null energy condition, we can claim that boundary scale invariance must lead to boundary conformal invariance. In our non-linear sigma model example, the target space metric must be strictly positive.
 Since any non-trivial field configuration that will violate the conformal invariance that are compatible with the gravity equation violates the strict null energy condition, the boundary scale invariance must imply boundary conformal invariance under the assumption of the strict null energy condition.

\section{Discussions}
In this paper, we have discussed boundary conditions for conformal field theories in general dimensions. We have found that Cardy's condition is necessary for the conformal invariance with unitarity, but it may not be sufficient in higher dimensions $d>3$. We conjecture that scale invariant boundary conditions will imply conformal invariant boundary conditions under a certain set of assumptions such as unitarity, causality, and discreteness of the spectrum. We do not have a non-perturbative proof from the field theory side. 

We have provided some evidence of the conjecture from a holographic approach. In the holographic approach, the boundary $g$-theorem, which is suitably extended to take into account the operator identities with a possible boundary virial current, enabled us to derive the conclusion that the trace of the boundary energy-momentum tensor must vanish at the scale invariant fixed point, and the conformal invariance of the boundary is obtained. We believe that the similar argument should exist in the field theory side.  It is important to understand a better formulation of the $g$-theorem from the field theory perspective.

We should note that although the strict null energy condition in holography is a sufficient condition for the unitarity of the boundary theories, boundary $g$-theorem and our proof of the enhancement of boundary conformal invariance from  boundary scale invariance, it may not be necessary and it is more desirable to provide more evidence. We have asked a similar question from the string compactication in \cite{Nakayama:2009fe}, but the analysis is restricted when there is no boundary or singularities, so it is interesting to revisit the question. In particular, our boundary condition allows ``negative tension" when $\eta^* < 0$, and it is important to understand whether it is reasonable or not.

Moreover, the null energy condition seems to be violated by quantum effects such as Casimir effects. We may be able to relax the condition by ``achronal averaged null energy condition" \cite{Graham:2007va} to avoid the Casimir energy violation. This seems more or less sufficient for our purposes because we have sufficiently large isometry in our study, but it has been later claimed even that version can be violated \cite{Urban:2009yt}. It will be a very interesting question whether that violation can be realized in our setup and can be fatal to our discussions.\footnote{Their example as it is does not invalidate our argument because their non-anomalous violation is time-dependent and it is against our assumption of Poincar\'e invariance, and their anomalous violation does not contribute to our setup due to the symmetry of the space-time.}

We may be able to extend our analysis to other defects in conformal field theories. One simple generalization is interfaces in conformal field theories. Indeed, the conformal interfaces can be not only regarded as a generalization of the boundaries, but can be seen as the same object by using the folding tricks to relate them to the boundaries of the tensor product conformal field theories. Once we assume the analogue of Cardy's condition for the interface, we may use the same trick for scale invariant interfaces and our conjecture should also apply. For instance it is interesting to understand whether the recently proposed renormalization group interface \cite{Gaiotto:2012np} preserves the conformal invariance.

It is not immediately clear whether the folding tricks work in  a holographic description, but certainly it seems possible to ``fill" the other side of the AdS space-time with the dual of the different bulk conformal field theory. The holographic dual of interface conformal field theories was pioneered in \cite{Yamaguchi:2002pa}, and the analogue of the $g$-theorem was proposed, so the direction is promising.

\section*{Acknowledgements}
We would like to thank A.~Stergiou and T.~Ugajin for discussions.
This work is supported by Sherman Fairchild Senior Research Fellowship at California Institute of Technology.

\end{document}